\documentclass[aps,pra,10pt,nofootinbib,superscriptaddress,twocolumn]{revtex4-2}
\usepackage[cm]{fullpage}
\usepackage{microtype}
\usepackage[T1]{fontenc}
\usepackage{times}
\usepackage{amsfonts,amsmath,amssymb,bbm,mathtools}
\newcommand{\maths}[1]{$\smash{#1}$}
\usepackage{graphicx,xcolor}
\usepackage[pdfstartview=FitH,colorlinks=true,citecolor=blue,linkcolor=blue,urlcolor=blue]{hyperref}

\begin{document}

\title{Critical velocity of a two-dimensional superflow past a potential barrier of arbitrary penetrability}

\author{Juliette Huynh}
\affiliation{Universit\'e C\^ote d'Azur, CNRS, INPHYNI, France}

\author{Fr\'ed\'eric H\'ebert}
\email{frederic.hebert@univ-cotedazur.fr}
\affiliation{Universit\'e C\^ote d'Azur, CNRS, INPHYNI, France}

\author{Mathias Albert}
\affiliation{Universit\'e C\^ote d'Azur, CNRS, INPHYNI, France}
\affiliation{Institut Universitaire de France (IUF)}

\author{Pierre-\'Elie Larr\'e}
\email{pierre-elie.larre@univ-cotedazur.fr}
\affiliation{Universit\'e C\^ote d'Azur, CNRS, INPHYNI, France}

\date{\today}

\begin{abstract}
We theoretically investigate the critical velocity for dissipationless motion of a two-dimensional superfluid past a static potential barrier of large width. The circular-shaped barrier provides a comprehensive analytical framework for the critical speed, for which we derive closed-form expressions using the hydraulic approximation, the hodograph method, and Janzen-Rayleigh expansions of the velocity potential. These analytical estimates are shown to be in good agreement with the numerical results of an imaginary-time integration of the full wave equation. In contrast to most of the state of the art, our study is not restricted to an impenetrable potential barrier nor to a quartic interaction Hamiltonian, which enables realistic modeling of recent experiments with atomic Bose-Einstein condensates and paraxial superfluids of light in two dimensions.
\end{abstract}

\maketitle

\section{Introduction}
\label{Sec:Introduction}

Superfluidity is the ability of a fluid to flow without dissipation~\cite{Leggett1999, Balibar2007, Pitaevskii2016}. Following its discovery in liquid helium-4~\cite{Kapitza1938, Allen1938}, Landau established that it should occur at flow speeds smaller than the critical velocity \maths{v_{\mathrm{c}}=\min_{\boldsymbol{p}}E_{\boldsymbol{p}}/p}, where \maths{E_{\boldsymbol{p}}} is the energy of an elementary excitation with momentum \maths{\boldsymbol{p}} in the fluid at rest~\cite{Landau1941a, Landau1941b}. However, Landau's criterion often overestimates the actual critical velocity for superfluidity because it does not properly account for the nonlinear nature of the interaction between the fluid and its environment~\cite{Feynman1957}. This is not only noticed in liquid helium-4 but also in many other systems evidencing a superfluid transition, which include liquid helium-3~\cite{Osheroff1972a, Osheroff1972b}, ultracold atomic vapors~\cite{Raman1999, Onofrio2000, Raman2001, Miller2007, Engels2007, Neely2010, Dries2010, Desbuquois2012, Kwon2015}, microcavity polariton condensates~\cite{Amo2009, Lerario2017}, and more recently paraxial superfluids of light~\cite{Vocke2016, Michel2018, Eloy2021}.

In the present work, we consider a two-dimensional (2D) superfluid flowing past a static potential barrier of large width. Drawing inspiration from the theoretical studies~\cite{Frisch1992, Rica2001, Pinsker2014} (see also Refs.~\cite{Josserand1997, Josserand1999, Huepe2000, Stiessberger2000, Sasaki2010, Singh2017, Pigeon2021, Stockdale2021, Muller2022, Kwak2023, Ronning2023} for related theoretical work), we analytically derive its critical speed using the hydraulic approximation, the hodograph method, and Janzen-Rayleigh expansions of the velocity potential. In contrast to previous work, we do it for a circular potential barrier of \textit{arbitrary} penetrability and a local interaction energy of \textit{arbitrary} dependence on the fluid density in the nonlinear Schr\"odinger equation. Comparison with imaginary-time numerical calculations inspired from the very recent study~\cite{Kwak2023} shows good agreement for different interaction potentials. To the best of our knowledge, no theoretical work had previously provided analytical results for the 2D critical velocity in the case of a penetrable potential barrier, which, however, has been recently measured in a quasi-2D atomic Bose-Einstein condensate~\cite{Kwon2015} without modeling other than numerical~\cite{Kwak2023}. In addition, no related work had previously analytically explored the possibility of a nonlinearity different from that of the usual nonlinear Schr\"odinger equation, thus providing theoretical support for different types of superfluids and, in particular, recent experiments with a 2D paraxial superfluid of light in a saturable nonlinear medium~\cite{Michel2018, Eloy2021}.

The article is organized as follows. In Sec.~\ref{Sec:HydrodynamicEquations}, we present the hydrodynamic equations of the superfluid, starting with the wave equation and two physical examples of it and finishing with the coupled system verified by the density and velocity fields. In Sec.~\ref{Sec:AnalyticalDerivationOfTheCriticalVelocity}, we then derive analytical results for the corresponding critical velocity for superfluidity. Using the hydraulic approximation and the hodograph method, the critical speed is obtained in the case of a penetrable and then impenetrable circular potential barrier by means of Janzen-Rayleigh expansions of the velocity potential around its incompressible approximation. Next, in Sec.~\ref{Sec:ComparisonWithNumericalSimulationsAndDiscussion}, we discuss these analytical results by confronting them to imaginary-time numerical simulations of the full wave equation. Finally, we conclude and give outlook to the present work in Sec.~\ref{Sec:ConclusionAndOutlook}. Details about the numerical calculation of the critical velocity are provided in the Appendix.

\section{Hydrodynamic equations}
\label{Sec:HydrodynamicEquations}

\subsection{Wave equation}
\label{SubSec:WaveEquation}

Many 2D nonlinear wave systems display superfluidity or an analogue of it when transported past obstacles (see Sec.~\ref{SubSec:PhysicalExamples} for two representative examples). These are often described by a complex-valued wave function \maths{\psi(\boldsymbol{r},t)} whose dependence on position \maths{\boldsymbol{r}=(x,y)=(r\cos\theta,r\sin\theta)} and time \maths{t} is governed by a nonlinear partial differential equation of the form
\begin{equation}
\label{Eq:GNLSE}
i\frac{\partial\psi}{\partial t}=-\frac{1}{2}\nabla^{2}\psi+U(\boldsymbol{r})\mathbbm{1}_{U}\psi+\epsilon(\rho)\psi,
\end{equation}
where \maths{\rho(\boldsymbol{r},t)=|\psi(\boldsymbol{r},t)|^{2}} is the density associated with \maths{\psi(\boldsymbol{r},t)}. In this generalization of the 2D nonlinear Schr\"odinger equation to an arbitrary local nonlinearity \maths{\epsilon(\rho)\psi} [the nonlinear Schr\"odinger equation is obtained for \maths{\epsilon(\rho)=\pm\rho}, i.e., for the quartic interaction Hamiltonian \maths{\pm\int d^{2}\boldsymbol{r}\,|\psi(\boldsymbol{r},t)|^{4}/2}], \maths{\nabla} denotes the del operator with respect to position and \maths{U(\boldsymbol{r})} is the potential of a static obstacle to the system's flow. This potential is assumed to be repulsive [\maths{U(\boldsymbol{r})>0}] and localized with typical width \maths{\sigma} [\maths{U(\boldsymbol{r})\to0} as \maths{r/\sigma\to\infty}], in such a way that the indicator \maths{\mathbbm{1}_{U}} equals \maths{1} or \maths{0} depending on whether \maths{U(\boldsymbol{r})} is penetrable or not, respectively [the condition for penetrability of \maths{U(\boldsymbol{r})} will be rigorously defined in Sec.~\ref{SubSec:PenetrableRegime}]. This distinction is based on the following heuristic reasoning. When \maths{U(\boldsymbol{r})} is penetrable, the superfluid occupies all the space and there is no reason to do anything about the potential barrier \maths{U(\boldsymbol{r})} in Eq.~\eqref{Eq:GNLSE}. When \maths{U(\boldsymbol{r})} is on the contrary impenetrable, the superfluid can only occupy the region \maths{r\gtrsim\sigma} where \maths{U(\boldsymbol{r})} is negligible, in such a way that the system's dynamics can be described by Eq.~\eqref{Eq:GNLSE} without any potential barrier \maths{U(\boldsymbol{r})} but with appropriate conditions for the wave function at the obstacle's boundary \maths{r\sim\sigma}. Finally, we assume that \maths{\epsilon(0)=0}, \maths{\epsilon(\rho)>0}, and \maths{(\partial\epsilon/\partial\rho)(\rho)>0} to prevent the system from developing modulational instabilities [e.g., \maths{\epsilon(\rho)=\rho} for the original nonlinear Schr\"odinger equation].

\subsection{Physical examples}
\label{SubSec:PhysicalExamples}

\subsubsection{Atomic Bose-Einstein condensates}
\label{SubSubSec:AtomicBoseEinsteinCondensates}

For example, Eq.~\eqref{Eq:GNLSE} is well known~\cite{Pitaevskii2016} to govern the dynamics of the 2D reduction \maths{\psi(\boldsymbol{r},t)} of the condensate wave function of an ultracold atomic Bose gas of three-dimensional (3D) \maths{\mathrm{s}}-wave scattering length \maths{a>0} in a tight one-dimensional (1D) harmonic potential \maths{\hbar^{2}z^{2}/(2m\ell^{4})}, where \maths{\hbar}, \maths{m}, and \maths{\ell} are respectively the reduced Planck constant, the atomic mass, and the harmonic length. In this 1D trap, the atoms almost live in the \maths{\boldsymbol{r}}-plane and interact with each other via a Hartree-Fock potential \maths{\epsilon(\rho)=g\rho^{\nu}} scaling as a positive power of the 2D number density \maths{\rho(\boldsymbol{r},t)=|\psi(\boldsymbol{r},t)|^{2}} when \maths{\rho a^{2}} is much smaller\footnote{But larger than \maths{(a/\ell)^{2}\exp[-(2\pi)^{1/2}\ell/a]} to prevent the gas from entering the 2D analogue of the Tonks-Girardeau regime. In this ultradilute regime, \maths{\epsilon(\rho)=4\pi(\hbar^{2}/m)\rho/|\!\ln(\rho\ell^{2})|} (see, e.g., Ref.~\cite{Petrov2000}).} or much larger than \maths{a/\ell}. In the first regime, \maths{mg/\hbar^{2}=(8\pi)^{1/2}a/\ell} and \maths{\nu=1}, and in the second one, \maths{\ell^{2/3}mg/\hbar^{2}=(3\pi/\sqrt{2})^{2/3}(a/\ell)^{2/3}} and \maths{\nu=2/3}~\cite{MunozMateo2008}. In this context, Eq.~\eqref{Eq:GNLSE} reads~\cite{Pitaevskii2016}
\begin{equation}
\label{Eq:Matter}
i\hbar\frac{\partial\psi}{\partial t}=-\frac{\hbar^{2}}{2m}\nabla^{2}\psi+U(\boldsymbol{r})\mathbbm{1}_{U}\psi+\epsilon(\rho)\psi,
\end{equation}
where the obstacle potential \maths{U(\boldsymbol{r})} is usually generated by means of a focused laser beam crossing the quasi-2D condensate perpendicularly~\cite{Desbuquois2012, Kwon2015}. We rigorously map Eq.~\eqref{Eq:Matter} to Eq.~\eqref{Eq:GNLSE} within the dimensionless variables \maths{\tilde{\boldsymbol{r}}=\boldsymbol{r}/\xi}, \maths{\tilde{t}=\mu t/\hbar}, \maths{\tilde{\psi}(\tilde{\boldsymbol{r}},\tilde{t})=\psi(\boldsymbol{r},t)/\bar{\rho}^{1/2}} [hence \maths{\tilde{\rho}(\tilde{\boldsymbol{r}},\tilde{t})=|\tilde{\psi}(\tilde{\boldsymbol{r}},\tilde{t})|^{2}=\rho(\boldsymbol{r},t)/\bar{\rho}}], \maths{\tilde{U}(\tilde{\boldsymbol{r}})=U(\boldsymbol{r})/\mu}, and
\begin{equation}
\label{Eq:MatterNL}
\tilde{\epsilon}(\tilde{\rho})=\frac{\epsilon(\rho)}{\mu}=\frac{\tilde{\rho}^{\nu}}{\nu},
\end{equation}
from which we eventually remove the tildes for readability. In these definitions, the proper units \maths{\xi=\hbar/(ms)} and \maths{\mu=ms^{2}} with
\begin{equation}
\label{Eq:MatterSound}
s=\sqrt{\frac{\bar{\rho}}{m}\frac{\partial\epsilon}{\partial\rho}(\bar{\rho})}=\sqrt{\frac{\nu g\bar{\rho}^{\nu}}{m}}
\end{equation}
are respectively the healing length, the chemical potential, and the speed of sound of the quasi-2D condensate at a typical number density \maths{\bar{\rho}} which we choose to be the uniform number density of the system in the absence of obstacle, i.e., when \maths{U(\boldsymbol{r})=0}.

\subsubsection{Paraxial superfluids of light}
\label{SubSubSec:ParaxialSuperfluidsOfLight}

Equation~\eqref{Eq:GNLSE} is also encountered in nonlinear optics to describe paraxial propagation of monochromatic light in nonlinear dielectrics~\cite{Boyd2020}. In this context, its resemblance with Eq.~\eqref{Eq:Matter} [see Eq.~\eqref{Eq:Light} below] has been used to investigate quantum hydrodynamic phenomena with classical nonlinear light~\cite{Coullet1989, Pomeau1993, Wan2007, Leboeuf2010}, leading to the research field of paraxial superfluids of light~\cite{Carusotto2014, Larre2015, Vocke2016, Michel2018, Fontaine2018, Rodrigues2020, Eloy2021}. Recent experiments~\cite{Michel2018, Eloy2021} have been done in a medium whose optical response to two lasers results in a refractive index of the form \maths{n_{0}+n_{1}(\boldsymbol{r})+n_{2}I/(1+I/I_{\mathrm{sat}})}. In this expansion, the middle contribution is defocusing [\maths{n_{1}(\boldsymbol{r})<0}] and induced by the first laser, of low intensity. The last one, which reproduces quite well the saturation of the optical nonlinearity~\cite{Kivshar2003} observed in photorefractive crystals~\cite{Michel2018, Eloy2021}, is also defocusing (\maths{n_{2}<0}) but induced by the second laser, of large intensity \maths{I(\boldsymbol{r},z)=n_{0}\varepsilon_{0}c\rho(\boldsymbol{r},z)/2}. In this equation, \maths{\varepsilon_{0}} and \maths{c} are the vacuum permittivity and speed of light, respectively, and \maths{\rho(\boldsymbol{r},z)=|\psi(\boldsymbol{r},z)|^{2}}, where \maths{\psi(\boldsymbol{r},z)} is the slowly varying envelope of the complex-valued electric field \maths{\psi(\boldsymbol{r},z)\exp[i(kz-\omega t)]} of the second laser, of angular frequency \maths{\omega} and carrier wave number \maths{k=n_{0}\omega/c} along the \maths{z}-axis. Defining \maths{U(\boldsymbol{r})=-(\omega/c)n_{1}(\boldsymbol{r})} and \maths{\epsilon(\rho)=-(\omega/c)n_{2}I/(1+I/I_{\mathrm{sat}})}, the equation for the electric-field amplitude \maths{\psi(\boldsymbol{r},z)} reads~\cite{Boyd2020}
\begin{equation}
\label{Eq:Light}
i\frac{\partial\psi}{\partial z}=-\frac{1}{2k}\nabla^{2}\psi+U(\boldsymbol{r})\mathbbm{1}_{U}\psi+\epsilon(\rho)\psi,
\end{equation}
whose formal analogy with Eq.~\eqref{Eq:Matter} is transparent. Equation~\eqref{Eq:Light} is cast into Eq.~\eqref{Eq:GNLSE} within the dimensionless variables defined in Sec.~\ref{SubSubSec:AtomicBoseEinsteinCondensates}, except that here \maths{\tilde{t}=\mu z},
\begin{equation}
\label{Eq:LightNL}
\tilde{\epsilon}(\tilde{\rho})=\bigg(1+\frac{1}{\tilde{\rho}_{\mathrm{sat}}}\bigg)^{2}\frac{\tilde{\rho}}{1+\tilde{\rho}/\tilde{\rho}_{\mathrm{sat}}},
\end{equation}
\maths{\xi=1/(ks)}, \maths{\mu=ks^{2}}, and
\begin{equation}
\label{Eq:LightSound}
s=\sqrt{\frac{\bar{\rho}}{k}\frac{\partial\epsilon}{\partial\rho}(\bar{\rho})}=\frac{\sqrt{|n_{2}|\varepsilon_{0}c\bar{\rho}/2}}{1+\bar{\rho}/\rho_{\mathrm{sat}}},
\end{equation}
where \maths{\tilde{\rho}_{\mathrm{sat}}=\rho_{\mathrm{sat}}/\bar{\rho}} with \maths{\rho_{\mathrm{sat}}=2I_{\mathrm{sat}}/(n_{0}\varepsilon_{0}c)}. Note that although~\eqref{Eq:LightSound} has no dimension here, we can call it ``speed of sound'' by analogy with Eq.~\eqref{Eq:MatterSound}.

\subsection{Superfluid hydrodynamics}
\label{SubSec:SuperfluidHydrodynamics}

We now express the wave function \maths{\psi(\boldsymbol{r},t)} in the polar form \maths{\psi(\boldsymbol{r},t)=\rho^{1/2}(\boldsymbol{r},t)\exp[i\phi(\boldsymbol{r},t)]} (the so-called Madelung representation), which inserted into Eq.~\eqref{Eq:GNLSE} leads to the usual hydrodynamic equations of atomic superfluids at zero temperature~\cite{Pitaevskii2016}:
\begin{equation}
\label{Eq:Hydro}
\begin{gathered}
\frac{\partial\rho}{\partial t}+\nabla\cdot(\rho\boldsymbol{v})=0, \\
\frac{\partial\boldsymbol{v}}{\partial t}+\nabla\bigg[\frac{v^{2}}{2}+U(\boldsymbol{r})\mathbbm{1}_{U}+\epsilon(\rho)-\frac{1}{2}\frac{\nabla^{2}\sqrt{\rho}}{\sqrt{\rho}}\bigg]=0,
\end{gathered}
\end{equation}
where \maths{\boldsymbol{v}(\boldsymbol{r},t)=\nabla\phi(\boldsymbol{r},t)} is the velocity field of the superfluid expressed in units of the speed of sound~\eqref{Eq:MatterSound} or~\eqref{Eq:LightSound} if one refers to the examples of superfluids given in Sec.~\ref{SubSec:PhysicalExamples}. The first of Eqs.~\eqref{Eq:Hydro} for the density \maths{\rho(\boldsymbol{r},t)} is nothing but the continuity equation while the second one for the velocity \maths{\boldsymbol{v}(\boldsymbol{r},t)} is Newton's second law of motion.

We are specifically interested in the solutions of Eqs.~\eqref{Eq:Hydro} that are typical of a superfluid flow, i.e., a flow which is (i) steady and (ii) devoid of any hydrodynamic disturbance far away from the obstacle~\cite{Frisch1992}: (i) \maths{\rho=\rho(\boldsymbol{r})} and \maths{\boldsymbol{v}=\boldsymbol{v}(\boldsymbol{r})}; (ii) \maths{\rho(\boldsymbol{r})=\rho_{\infty}=\mathrm{const}} and \maths{\boldsymbol{v}(\boldsymbol{r})=\boldsymbol{v}_{\infty}=\boldsymbol{\mathrm{const}}} at infinity where \maths{U(\boldsymbol{r})=0}, and we choose \maths{\rho_{\infty}=1} in adequacy with the definition of the dimensioned density \maths{\bar{\rho}} given in Sec.~\ref{SubSec:PhysicalExamples}, as well as \maths{\boldsymbol{v}_{\infty}=(v_{\infty},0)} with \maths{v_{\infty}>0} (asymptotic flow from left to right) for the sake of concreteness. As such, \maths{\rho(\boldsymbol{r})} and \maths{\boldsymbol{v}(\boldsymbol{r})} verify the following differential system:
\begin{equation}
\label{Eq:HydroSF}
\begin{gathered}
\nabla\cdot(\rho\boldsymbol{v})=0, \\
\frac{v^{2}}{2}+U(\boldsymbol{r})\mathbbm{1}_{U}+\epsilon(\rho)-\frac{1}{2}\frac{\nabla^{2}\sqrt{\rho}}{\sqrt{\rho}}=\frac{v_{\infty}^{2}}{2}+\epsilon(1),
\end{gathered}
\end{equation}
from which one sees that the condition of existence of its solutions bears on \maths{v_{\infty}} once the potentials \maths{U(\boldsymbol{r})} and \maths{\epsilon(\rho)} are fixed. It is this constraint on \maths{v_{\infty}} we seek to determine in the present work. We will show that to have superfluidity, \maths{v_{\infty}} must be smaller than a critical speed \maths{v_{\mathrm{c}}} specific to the \maths{U(\boldsymbol{r})} and \maths{\epsilon(\rho)} considered. Analytical results for this critical velocity for superfluid motion are derived in Sec.~\ref{Sec:AnalyticalDerivationOfTheCriticalVelocity} and confronted to numerical calculations in Sec.~\ref{Sec:ComparisonWithNumericalSimulationsAndDiscussion}.

It is worth noting that when \maths{U(\boldsymbol{r})\to0}, linear-response theory applies\footnote{Except in the vicinity of the origin (\maths{r\to0}) and close to the sound barrier (\maths{v_{\infty}\to1}) where the density response function diverges logarithmically and algebraically, respectively.} and predicts that \maths{v_{\mathrm{c}}} equals unity~\cite{Hakim1997, Pavloff2002, Astrakharchik2004}, in agreement with Landau's criterion \maths{v_{\mathrm{c}}=\min_{\boldsymbol{p}}E_{\boldsymbol{p}}/p}, where \maths{E_{\boldsymbol{p}}=p(1+p^{2}/4)^{1/2}} is the Bogoliubov dispersion relation of the superfluid far away from the obstacle and in the comoving frame~\cite{Pitaevskii2016}. Thus, for an obstacle potential \maths{U(\boldsymbol{r})} with arbitrary (so possibly large) amplitude, \maths{v_{\mathrm{c}}} must necessarily be smaller than unity and we consequently restrict our study to the subsonic regime \maths{v_{\infty}<1}.

\section{Analytical derivation of the critical velocity}
\label{Sec:AnalyticalDerivationOfTheCriticalVelocity}

\subsection{Hydraulic approximation}
\label{SubSec:HydraulicApproximation}

From now on, we consider that the typical width of the potential barrier \maths{U(\boldsymbol{r})} is very large: \maths{\sigma\to\infty}. In this case, the typical scale of variation of the superfluid density \maths{\rho(\boldsymbol{r})} is of the order of \maths{\sigma} (see, e.g., Ref.~\cite{Hakim1997} although in 1D). As a result, one can neglect the dispersive term \maths{-\nabla^{2}\rho^{1/2}/(2\rho^{1/2})} in the second of Eqs.~\eqref{Eq:HydroSF}, which thus simplifies to an algebraic equation for the density \maths{\rho} as a function of the norm \maths{v=|\nabla\phi|} of the velocity. In this hydraulic~\cite{Hakim1997} approach, we are left with the following differential problem for the velocity potential \maths{\phi(\boldsymbol{r})}:
\begin{equation}
\label{Eq:HydroSFHydrau}
\begin{gathered}
\nabla\cdot[\rho(|\nabla\phi|)\nabla\phi]=0, \\
\epsilon[\rho(v)]=\epsilon(1)-U(\boldsymbol{r})\mathbbm{1}_{U}-\frac{v^{2}-v_{\infty}^{2}}{2},
\end{gathered}
\end{equation}
with
\begin{equation}
\label{Eq:AC}
\phi|_{r/\sigma\to\infty}=v_{\infty}x=v_{\infty}r\cos\theta
\end{equation}
as asymptotic condition.

Note that the analytical treatment of the so-called quantum pressure \maths{-\nabla^{2}\rho^{1/2}/(2\rho^{1/2})} is difficult in 2D, especially when \maths{\sigma} is small (see discussions in Secs.~\ref{Sec:ComparisonWithNumericalSimulationsAndDiscussion} and~\ref{Sec:ConclusionAndOutlook}). As far as we know, the only work in which this issue is tackled is Ref.~\cite{Pinsker2014}, where the critical velocity for superfluid motion past an impenetrable potential barrier is perturbatively estimated up to first order in \maths{1/\sigma^{2}\to0}. This result does not add much to the hydraulic physics and we consequently restrict our study to the approximation explained above, as in most of the literature dealing with the superfluid transition in 2D, starting with the seminal work~\cite{Frisch1992}.

\subsection{Hodograph method}
\label{SubSec:HodographMethod}

Finding the condition of existence of the superflow described by Eqs.~\eqref{Eq:HydroSFHydrau} and~\eqref{Eq:AC} is facilitated in the hodograph plane~\cite{Landau1987} where the first of Eqs.~\eqref{Eq:HydroSFHydrau}, nonlinear, is transformed into the
following linear equation:
\begin{equation}
\label{Eq:Hodograph}
\rho(v)v^{2}\frac{\partial^{2}\Phi}{\partial v^{2}}+\frac{\partial}{\partial v}[\rho(v)v]v\frac{\partial\Phi}{\partial v}+\frac{\partial}{\partial v}[\rho(v)v]\frac{\partial^{2}\Phi}{\partial\vartheta^{2}}=0.
\end{equation}
In this equation, \maths{\Phi(\boldsymbol{v})=\boldsymbol{v}\cdot\boldsymbol{r}-\phi(\boldsymbol{r})} is the hodograph transform of \maths{\phi(\boldsymbol{r})} and \maths{\vartheta} is the angular coordinate of \maths{\boldsymbol{v}=(v_{x},v_{y})=(v\cos\vartheta,v\sin\vartheta)} in polar representation. Focusing on the equation of the characteristic curves~\cite{Landau1987} of Eq.~\eqref{Eq:Hodograph}:
\begin{equation}
\label{Eq:Characteristics}
\frac{\partial}{\partial v}[\rho(v)v]dv^{2}+\rho(v)v^{2}d\vartheta^{2}=0,
\end{equation}
one then infers that there is no trajectory \maths{\vartheta=\vartheta(v)} along which a possible wave discontinuity can propagate---a hallmark of superfluid motion---provided
\begin{equation}
\label{Eq:SF}
\frac{\partial}{\partial v}[\rho(v)v]>0\quad\forall v,
\end{equation}
which is the constraint for \maths{d\vartheta/dv} to be complex-valued, i.e., for Eq.~\eqref{Eq:Hodograph} to be elliptic. This condition for superfluidity has long been used to investigate the superfluid transition in 2D, starting with Ref.~\cite{Frisch1992}. Nevertheless, as far as we know, the reasoning leading to it has never really been made explicit in the superfluid literature, what we have tried to overcome in the present paragraph.

Using the identity \maths{\partial\rho/\partial v=(\partial\epsilon/\partial v)/(\partial\epsilon/\partial\rho)} and the second of Eqs.~\eqref{Eq:HydroSFHydrau}, it is easy to show that the left-hand side of inequality~\eqref{Eq:SF} equals \maths{\rho(\boldsymbol{r})[1-v^{2}(\boldsymbol{r})/s^{2}(\boldsymbol{r})]}, where \maths{s(\boldsymbol{r})=\{\rho(\boldsymbol{r})(\partial\epsilon/\partial\rho)[\rho(\boldsymbol{r})]\}^{1/2}} is the local speed of sound.\footnote{Reformulated in these terms, Eq.~\eqref{Eq:Hodograph} is identical to Chaplygin's equation of gas dynamics~\cite{Landau1987}.} Thus, condition~\eqref{Eq:SF} is equivalent to \maths{v^{2}(\boldsymbol{r})<s^{2}(\boldsymbol{r})} for all \maths{\boldsymbol{r}}, which is the local Landau criterion for superfluidity~\cite{Hakim1997} after removing the squares. This constraint is also equivalent to the same inequality with \maths{v(\boldsymbol{r})} and \maths{s(\boldsymbol{r})} respectively replaced with their maximum \maths{v_{\mathrm{max}}} and minimum \maths{s_{\mathrm{min}}}, the latter being reached at the minimum density \maths{\rho_{\mathrm{min}}} given the interaction potentials \maths{\epsilon(\rho)} considered in this work. By relating \maths{\rho_{\mathrm{min}}} to \maths{v_{\mathrm{max}}} and \maths{v_{\infty}} using the second of Eqs.~\eqref{Eq:HydroSFHydrau} with \maths{U(\boldsymbol{r})} replaced with its maximum \maths{U_{\mathrm{max}}} since \maths{U(\boldsymbol{r})} is repulsive, we thus come to the following superfluid condition in terms of \maths{v_{\mathrm{max}}}, \maths{v_{\infty}}, and \maths{U_{\mathrm{max}}}:
\begin{equation}
\label{Eq:LL}
\begin{gathered}
v_{\mathrm{max}}^{2}<\rho_{\mathrm{min}}^{\vphantom{2}}\frac{\partial\epsilon}{\partial\rho}(\rho_{\mathrm{min}}^{\vphantom{2}}), \\
\epsilon(\rho_{\mathrm{min}})=\epsilon(1)-U_{\mathrm{max}}\mathbbm{1}_{U}-\frac{v_{\mathrm{max}}^{2}-v_{\infty}^{2}}{2}.
\end{gathered}
\end{equation}

Constraint~\eqref{Eq:LL} changes from one interaction potential \maths{\epsilon(\rho)} to another. Hereafter, we explicit it in the case where \maths{\epsilon(\rho)} is given by Eq.~\eqref{Eq:MatterNL}:
\begin{equation}
\label{Eq:MatterLL}
\Big(1+\frac{\nu}{2}\Big)v_{\mathrm{max}}^{2}-\frac{\nu}{2}v_{\infty}^{2}<1-\nu U_{\mathrm{max}}^{\vphantom{2}}\mathbbm{1}_{U},
\end{equation}
which is, for \maths{\nu=1} and \maths{\mathbbm{1}_{U}=0}, the condition for superfluidity first established in Ref.~\cite{Frisch1992}. An explicit expression for~\eqref{Eq:LL} also exists in the case where \maths{\epsilon(\rho)} is given by Eq.~\eqref{Eq:LightNL} but it is cumbersome.

\subsection{Model obstacle}
\label{SubSec:ModelObstacle}

In order to find the critical velocity for superfluidity \maths{v_{\mathrm{c}}}, one needs to relate \maths{v_{\mathrm{max}}=\max_{\boldsymbol{r}}|\nabla\phi(\boldsymbol{r})|} to \maths{v_{\infty}} in~\eqref{Eq:LL}, which requires to solve the first of Eqs.~\eqref{Eq:HydroSFHydrau} for \maths{\phi(\boldsymbol{r})}. The procedure obviously depends on the shape of the obstacle, which we choose to be represented by the circular potential barrier
\begin{equation}
\label{Eq:Obstacle}
U(\boldsymbol{r})=
\begin{dcases}
U_{0}>0 & \text{if \maths{r<\sigma}} \\
0 & \text{otherwise}
\end{dcases}
,
\end{equation}
hence \maths{U_{\mathrm{max}}=U_{0}} in~\eqref{Eq:LL}. Given the central symmetry of~\eqref{Eq:Obstacle}, Eqs.~\eqref{Eq:HydroSFHydrau} and~\eqref{Eq:AC} will be naturally analyzed in the polar coordinates \maths{r} and \maths{\theta}.

It is worth noting that a potential barrier of the form \maths{U(\boldsymbol{r})=U_{0}\exp(-r^{2}/\sigma^{2})} would have been a better choice for comparison with experiments with atomic Bose-Einstein condensates or paraxial superfluids of light (see Sec.~\ref{SubSec:PhysicalExamples}), where obstacles are often Gaussian-shaped. However, this potential does not make it possible to push analytics further than Eqs.~\eqref{Eq:LL}, which is not the case of the potential~\eqref{Eq:Obstacle}, and it has already been numerically investigated in a very recent work~\cite{Kwak2023}. In addition, by reproducing the results of this reference, we have noticed that they do not differ much from those obtained for~\eqref{Eq:Obstacle} (see comments in Sec.~\ref{Sec:ComparisonWithNumericalSimulationsAndDiscussion} and the Appendix). Thus, the simple model~\eqref{Eq:Obstacle} turns out to be well suited to describe realistic situations while providing a comprehensive analytical framework for the 2D critical speed.

\subsection{Penetrable regime}
\label{SubSec:PenetrableRegime}

When~\eqref{Eq:Obstacle} is penetrable (\maths{\mathbbm{1}_{U}=1}), we close Eqs.~\eqref{Eq:HydroSFHydrau} and~\eqref{Eq:AC} with the following continuity equations at the boundary \maths{r=\sigma}:
\begin{equation}
\label{Eq:BCPenetrable}
\begin{gathered}
\phi|_{r=\sigma^{-}}=\phi|_{r=\sigma^{+}}, \\
\rho(|\nabla\phi|)\frac{\partial\phi}{\partial r}\bigg|_{r=\sigma^{-}}=\rho(|\nabla\phi|)\frac{\partial\phi}{\partial r}\bigg|_{r=\sigma^{+}}.
\end{gathered}
\end{equation}
While the first of Eqs.~\eqref{Eq:BCPenetrable} imposes no phase jump for the superfluid wave function at \maths{r=\sigma}, the second equation for the radial component of the current density \maths{\rho(\boldsymbol{r})\nabla\phi(\boldsymbol{r})} follows from the first of Eqs.~\eqref{Eq:HydroSFHydrau} integrated along an arbitrary radial cut of a thin annulus of median radius \maths{\sigma\to\infty}.

\subsubsection{Incompressible approximation}
\label{SubSubSec:IncompressibleApproximation}

We start by solving Eqs.~\eqref{Eq:HydroSFHydrau},~\eqref{Eq:AC}, and~\eqref{Eq:BCPenetrable} by neglecting
\begin{equation}
\label{Eq:Incompressible}
\frac{v^{2}(\boldsymbol{r})-v_{\infty}^{2}}{2}=\frac{\chi=v_{\infty}^{2}}{2}\bigg[\frac{|\nabla\phi(\boldsymbol{r})|^{2}}{v_{\infty}^{2}}-1\bigg]\xrightarrow[\chi\to0]{}0
\end{equation}
in the right-hand side of the second of Eqs.~\eqref{Eq:HydroSFHydrau}. This can be seen as an incompressible approximation for the superfluid since the dimensionless parameter \maths{\chi} defined in~\eqref{Eq:Incompressible} is also expressed in terms of the dimensioned quantities of Sec.~\ref{SubSec:PhysicalExamples} as \maths{\chi=v_{\infty}^{2}/s^{2}=(m\text{ or }k)v_{\infty}^{2}\kappa}, where \maths{\kappa} denotes the compressibility of the superfluid at the uniform density \maths{\bar{\rho}}~\cite{Pitaevskii2016}.

\begin{figure*}[t!]
\centering
\includegraphics[scale=0.9]{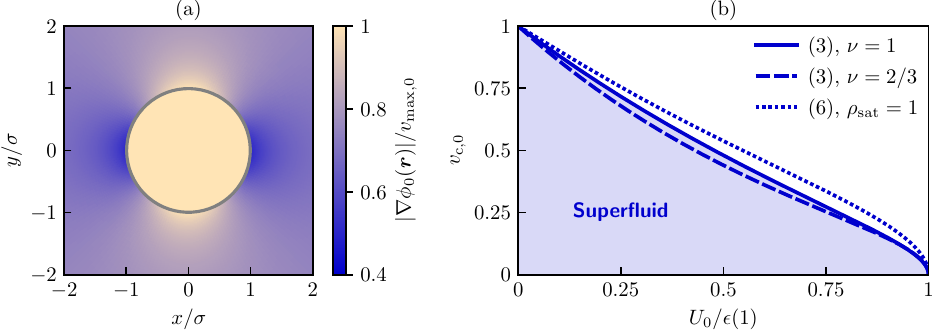}
\caption{(a) Colormap of the norm of the velocity field \maths{\nabla\phi_{0}(\boldsymbol{r})} normalized to its maximum value \maths{v_{\mathrm{max},0}} in the incompressible approximation \maths{\chi=0}, obtained from Eq.~\eqref{Eq:Phi0} for the interaction potential~\eqref{Eq:MatterNL} with \maths{\nu=1} and for the potential barrier~\eqref{Eq:Obstacle} with \maths{U_{0}/\epsilon(1)=0.5}, the boundary of which is materialized by the circle of proper radius \maths{\sigma}. (b) Critical velocity for superfluidity \maths{v_{\mathrm{c},0}} in the incompressible approximation \maths{\chi=0} and in the penetrable regime \maths{U_{0}/\epsilon(1)<1} when \maths{\epsilon(\rho)} is given by Eq.~\eqref{Eq:MatterNL} with \maths{\nu=1} (solid curve), by Eq.~\eqref{Eq:MatterNL} with \maths{\nu=2/3} (dashed curve), and by Eq.~\eqref{Eq:LightNL} with \maths{\rho_{\mathrm{sat}}=1} (dotted curve). For each curve, any point \maths{(U_{0},v_{\infty})} below corresponds to a superfluid flow (shaded region below the solid curve for instance).}
\label{Fig:Vc0}
\end{figure*}

In this approximation, the density \maths{\rho(\boldsymbol{r})=\rho_{0}(\boldsymbol{r})} is constant on either side of the obstacle's boundary \maths{r=\sigma}:
\begin{equation}
\label{Eq:PiecewiseConst}
\rho_{0}(\boldsymbol{r})=
\begin{dcases}
\rho_{0}=\epsilon^{-1}[\epsilon(1)-U_{0}] & \text{if \maths{r<\sigma}} \\
1 & \text{otherwise}
\end{dcases}
,
\end{equation}
which makes the first of Eqs.~\eqref{Eq:HydroSFHydrau} simplify to the following 2D Laplace equation for \maths{\phi(\boldsymbol{r})=\phi_{0}(\boldsymbol{r})}:
\begin{equation}
\label{Eq:Laplace}
\nabla^{2}\phi_{0}=0\quad\forall r\lessgtr\sigma.
\end{equation}
It is worth noting that given Eq.~\eqref{Eq:PiecewiseConst}, the positiveness of \maths{\epsilon(\rho)} imposes
\begin{equation}
\label{Eq:Penetrable}
U_{0}<\epsilon(1),
\end{equation}
which can be considered as the condition for penetrability of the obstacle potential~\eqref{Eq:Obstacle}---one thus has \maths{\mathbbm{1}_{U}=H[\epsilon(1)-U_{0}]}, where \maths{H} is the Heaviside step function. Given the general form of the solution of Eq.~\eqref{Eq:Laplace} in polar coordinates, which is \maths{\phi_{0}(r,\theta)=a_{0}+b_{0}\ln r+\sum_{k=1}^{\infty}R_{k}(r)\Theta_{k}(\theta)} with \maths{R_{k}(r)=a_{k}r^{k}+b_{k}r^{-k}}, \maths{\Theta_{k}(\theta)=c_{k}\cos(k\theta)+d_{k}\sin(k\theta)}, and \maths{a_{0,k},b_{0,k},c_{k},d_{k}=\mathrm{const}}, solving Eq.~\eqref{Eq:Laplace} with the asymptotic condition~\eqref{Eq:AC} and the boundary conditions~\eqref{Eq:BCPenetrable} becomes an easy task. One finds
\begin{equation}
\label{Eq:Phi0}
\phi_{0}(\boldsymbol{r})=v_{\infty}r\cos\theta\times
\begin{dcases}
\frac{2}{1+\rho_{0}} & \text{if \maths{r<\sigma}} \\
1+\frac{1-\rho_{0}}{1+\rho_{0}}\frac{\sigma^{2}}{r^{2}} & \text{otherwise}
\end{dcases}
,
\end{equation}
from which one infers that \maths{v_{\mathrm{max},0}=\max_{\boldsymbol{r}}|\nabla\phi_{0}(\boldsymbol{r})|} is reached everywhere in the disk of radius \maths{\sigma} [see Fig.~\ref{Fig:Vc0}(a) for visualization] and reads
\begin{equation}
\label{Eq:Vmax0}
v_{\mathrm{max},0}=\frac{2}{1+\rho_{0}}v_{\infty},
\end{equation}
where \maths{\rho_{0}}, the superfluid density inside the obstacle potential, is a function of \maths{U_{0}} defined in Eq.~\eqref{Eq:PiecewiseConst}. Inserting~\eqref{Eq:Vmax0} into~\eqref{Eq:LL} yields an inequality on \maths{v_{\infty}} and \maths{U_{0}} which can be rearranged in the form \maths{v_{\infty}<v_{\mathrm{c},0}}, where the \maths{U_{0}}-dependent velocity \maths{v_{\mathrm{c},0}} is the critical speed we are looking for.

For example, when \maths{\epsilon(\rho)} is given by Eq.~\eqref{Eq:MatterNL}, one has \maths{\rho_{0}=(1-\nu U_{0})^{1/\nu}}, \maths{v_{\mathrm{max},0}=2v_{\infty}/[1+(1-\nu U_{0})^{1/\nu}]}, and inequality~\eqref{Eq:MatterLL} becomes \maths{v_{\infty}<v_{\mathrm{c},0}} with
\begin{equation}
\label{Eq:MatterVc0Penetrable}
v_{\mathrm{c},0}=\sqrt{\frac{1-\nu U_{0}}{4(1+\nu/2)/[1+(1-\nu U_{0})^{1/\nu}]^{2}-\nu/2}}.
\end{equation}
A closed-form but unhandy expression for \maths{v_{\mathrm{c},0}} also exists in the case where \maths{\epsilon(\rho)} is given by Eq.~\eqref{Eq:LightNL}. In Fig.~\ref{Fig:Vc0}(b), we plot these \maths{v_{\mathrm{c},0}}'s as a function of \maths{U_{0}/\epsilon(1)<1} for \maths{\nu=1} and \maths{2/3} when \maths{\epsilon(\rho)} is given by Eq.~\eqref{Eq:MatterNL} and for \maths{\rho_{\mathrm{sat}}=1} when \maths{\epsilon(\rho)} is given by Eq.~\eqref{Eq:LightNL}. As expected, all curves converge to Landau's critical speed when \maths{U_{0}/\epsilon(1)\to0}. On the other hand, they all drop to zero when \maths{U_{0}/\epsilon(1)\to1}, which can be explained as follows. In the penetrable regime \maths{U_{0}/\epsilon(1)<1}, the maximum of the norm of the velocity field is reached and thus the superfluid transition takes place inside the obstacle potential where the superfluid density drops to zero when \maths{U_{0}/\epsilon(1)\to1}, so does the corresponding local speed of sound. Since this speed of sound is an upper bound for the critical velocity for superfluidity, it is then normal for the latter to vanish in this limit. However, superfluidity is not irretrievably lost from entry to the impenetrable regime \maths{U_{0}/\epsilon(1)>1} because the fluid can always go around the obstacle from the north and the south (see Sec.~\ref{SubSec:ImpenetrableRegime}). This contrasts with the 1D geometry (see, e.g.,~\cite{Huynh2022} and references therein) where the fluid is, in this impenetrable regime and in the hydraulic approximation, cut into two disconnected parts. Coming back to the present situation, the crossing from below of the critical frontiers displayed in Fig.~\ref{Fig:Vc0}(b) is typically marked by the nucleation of a rarefaction wave~\cite{Pinsker2014, Kwak2023} known as Jones-Roberts soliton~\cite{Jones1982, Meyer2017} inside the obstacle.\footnote{This is maybe what is observed in Figs.~2e and~2f of Ref.~\cite{Eloy2021}.}

\subsubsection{Janzen-Rayleigh expansions}
\label{SubSubSec:JanzenRayleighExpansions}

It is possible to refine the results of Sec.~\ref{SubSubSec:IncompressibleApproximation}, valid in the incompressible approximation \maths{\chi=0}, by perturbatively treating the velocity term~\eqref{Eq:Incompressible} in the second of Eqs.~\eqref{Eq:HydroSFHydrau}. To do so, we draw inspiration from the seminal work~\cite{Rica2001} by searching for the velocity potential in the Janzen-Rayleigh form~\cite{Janzen1913, Rayleigh1916}
\begin{equation}
\label{Eq:JR}
\phi(\boldsymbol{r})=\sum_{k=0}^{n}\phi_{k}(\boldsymbol{r})\chi^{k}+o(\chi^{n}),
\end{equation}
where \maths{\phi_{0}(\boldsymbol{r})} is given in Eq.~\eqref{Eq:Phi0}, \maths{\chi} tends to zero, and \maths{n\geqslant0} is the order of the expansion in powers of \maths{\chi}. Solving Eqs.~\eqref{Eq:HydroSFHydrau},~\eqref{Eq:AC}, and~\eqref{Eq:BCPenetrable} order by order using Eq.~\eqref{Eq:JR}, one obtains \maths{v_{\mathrm{max}}} in the form
\begin{equation}
\label{Eq:JRVmax}
v_{\mathrm{max}}=\sum_{k=0}^{n}v_{\mathrm{max},k}\chi^{k}+o(\chi^{n}),
\end{equation}
where \maths{v_{\mathrm{max},0}} is given in Eq.~\eqref{Eq:Vmax0} and the other \maths{v_{\mathrm{max},k}}'s are functions of \maths{v_{\infty}} and \maths{U_{0}} as complex as \maths{k} is large. Expanding~\eqref{Eq:LL} up to order \maths{n} in \maths{\chi} using~\eqref{Eq:JRVmax} and rewriting \maths{\chi} as \maths{v_{\infty}^{2}}, one eventually gets~\eqref{Eq:LL} as a constraint on \maths{v_{\infty}} and \maths{U_{0}} only, which can be in principle expressed in the form \maths{v_{\infty}<v_{\mathrm{c},n}} with \maths{v_{\mathrm{c},n}=v_{\mathrm{c},n}(U_{0})} being the critical velocity for superfluidity at order \maths{n} in the Janzen-Rayleigh expansion~\eqref{Eq:JR}.

\begin{table}[t!]
\centering
\begin{tabular}{c@{\quad}c@{\quad}c@{\quad}c}
\hline\hline
\maths{\epsilon(\rho)} & \eqref{Eq:MatterNL}, \maths{\nu=1} & \eqref{Eq:MatterNL}, \maths{\nu=2/3} & \eqref{Eq:LightNL}, \maths{\rho_{\mathrm{sat}}=1} \\
\maths{v_{\mathrm{c},0}} for \maths{U_{0}=\epsilon(1)/2} & 0.48(0) & 0.44(0) & 0.53(8) \\
\maths{v_{\mathrm{c},1}} for \maths{U_{0}=\epsilon(1)/2} & 0.44(2) & 0.40(4) & 0.48(8) \\
\maths{v_{\mathrm{c},2}} for \maths{U_{0}=\epsilon(1)/2} & 0.43(7) & 0.40(0) & 0.48(2) \\
\hline\hline
\end{tabular}
\caption{Proof by example of the accuracy of the Janzen-Rayleigh method in determining the critical velocity for superfluidity beyond the incompressible approximation \maths{\chi=0} (\maths{v_{\mathrm{c},0}}), here in the penetrable regime \maths{U_{0}<\epsilon(1)}: Between orders \maths{n=1} and \maths{2}, the critical speed \maths{v_{\mathrm{c}}} barely varies by \maths{(1-2)\%}; From left to right, one has \maths{1-v_{\mathrm{c},2}/v_{\mathrm{c},1}\simeq(\text{\maths{1.13}, \maths{0.99}, and \maths{1.23}})\times10^{-2}}.}
\label{Tab:VcPenetrable}
\end{table}

For the three interaction potentials \maths{\epsilon(\rho)} considered in Fig.~\ref{Fig:Vc0}(b), a relative difference of \maths{(1-2)\%} is observed for the critical speed between orders \maths{n=1} and \maths{2} of the Janzen-Rayleigh expansion~\eqref{Eq:JR}, as Tab.~\ref{Tab:VcPenetrable} shows for the median obstacle amplitude \maths{U_{0}=\epsilon(1)/2}. To be quantitative, we provide below the recurrence relations between the \maths{\phi_{k}(\boldsymbol{r})}'s of the Janzen-Rayleigh expansion~\eqref{Eq:JR} when \maths{\epsilon(\rho)} is given by Eq.~\eqref{Eq:MatterNL}:
\begin{align}
\notag
(1-\nu U_{0})\nabla^{2}\phi_{k+1}&\left.=\frac{1}{2v_{\infty}^{2}}\sum_{j=0}^{k}\Big[\nabla\phi_{k-j}\cdot\nabla (v^{2})_{j}\right. \\
\label{Eq:MatterJRPenetrable1}
&\left.\hphantom{=}+\nu\nabla^{2}\phi_{k-j}(v^{2})_{j}\Big]-\frac{\nu}{2}\nabla^{2}\phi_{k},\right.
\end{align}
\begin{equation}
\label{Eq:MatterJRPenetrable2}
\phi_{k}|_{r=\sigma^{-}}=\phi_{k}|_{r=\sigma^{+}},
\end{equation}
\begin{align}
\notag
&\sum_{h,i,j=0}^{\infty}\bigg({-}\frac{\nu}{2}\bigg)^{h}\binom{1/\nu}{h}\binom{h}{i}\frac{i!}{\prod_{\ell=1}^{\infty}i_{\ell}!}\bigg\{(1-\nu U_{0})^{1/\nu-h} \\
\notag
&\times\bigg[((v^{2})_{0}-1)^{h}\prod_{\ell=1}^{\infty}\bigg(\frac{(v^{2})_{\ell}}{(v^{2})_{0}-1}\bigg)^{i_{\ell}}\frac{\partial\phi_{j}}{\partial r}\bigg]_{r=\sigma^{-}} \\
\label{Eq:MatterJRPenetrable3}
&-\bigg[((v^{2})_{0}-1)^{h}\prod_{\ell=1}^{\infty}\bigg(\frac{(v^{2})_{\ell}}{(v^{2})_{0}-1}\bigg)^{i_{\ell}}\frac{\partial\phi_{j}}{\partial r}\bigg]_{r=\sigma^{+}}\bigg\}=0.
\end{align}
In these equations, \maths{(v^{2})_{k}=\sum_{j=0}^{k}\nabla\phi_{k-j}\cdot\nabla\phi_{j}} and the indices \maths{h}, \maths{i}, and \maths{j} of the summation in Eq.~\eqref{Eq:MatterJRPenetrable3} are such that \maths{h+\sum_{\ell=1}^{\infty}\ell i_{\ell}+j=k} with \maths{\sum_{\ell=1}^{\infty}i_{\ell}=i}. Analytical formulas also exist when \maths{\epsilon(\rho)} is given by Eq.~\eqref{Eq:LightNL} but their very long length makes their writing unreasonable.

\subsection{Impenetrable regime}
\label{SubSec:ImpenetrableRegime}

When~\eqref{Eq:Obstacle} is impenetrable [\maths{U_{0}>\epsilon(1)} and so \maths{\mathbbm{1}_{U}=0}], the superfluid only occupies the region \maths{r>\sigma} and we just need to supplement Eq.~\eqref{Eq:AC} with one boundary condition to close the differential problem for \maths{\phi(\boldsymbol{r})}. We choose this condition in the form
\begin{equation}
\label{Eq:BCImpenetrable}
\frac{\partial\phi}{\partial r}\bigg|_{r=\sigma}=0,
\end{equation}
which corresponds to the usual no-slip boundary condition of classical hydrodynamics: The flow velocity at the boundary of a rigid body is tangential~\cite{Landau1987}. Interestingly, Eqs.~\eqref{Eq:HydroSFHydrau},~\eqref{Eq:AC}, and~\eqref{Eq:BCImpenetrable}, which constitute the differential problem investigated in, e.g., Refs.~\cite{Frisch1992, Rica2001, Pinsker2014}, do not involve \maths{U_{0}}, in such a way that the corresponding critical velocity for superfluidity does not depend on this parameter.

\begin{table}[t!]
\centering
\begin{tabular}{c@{\quad}c@{\quad}c@{\quad}c}
\hline\hline
\maths{\epsilon(\rho)} & \eqref{Eq:MatterNL}, \maths{\nu=1} & \eqref{Eq:MatterNL}, \maths{\nu=2/3} & \eqref{Eq:LightNL}, \maths{\rho_{\mathrm{sat}}=1} \\
\maths{v_{\mathrm{c},0}} & 0.42(6) & 0.44(7) & 0.49(1) \\
\maths{v_{\mathrm{c},1}} & 0.39(0) & 0.40(7) & 0.44(2) \\
\maths{v_{\mathrm{c},2}} & 0.38(0) & 0.39(6) & 0.42(9) \\
\maths{v_{\mathrm{c},3}} & 0.37(5) & 0.39(1) & 0.42(3) \\
\hline\hline
\end{tabular}
\caption{Same as Tab.~\ref{Tab:VcPenetrable} but in the impenetrable regime \maths{U_{0}>\epsilon(1)}. A relative difference of barely \maths{(1-2)\%} is observed for \maths{v_{\mathrm{c}}} between orders \maths{n=2} and \maths{3}; From left to right, one has \maths{1-v_{\mathrm{c},3}/v_{\mathrm{c},2}\simeq(\text{\maths{1.32}, \maths{1.26}, and \maths{1.40}})\times10^{-2}}.}
\label{Tab:VcImpenetrable}
\end{table}

\begin{figure*}[t!]
\centering
\includegraphics[scale=0.9]{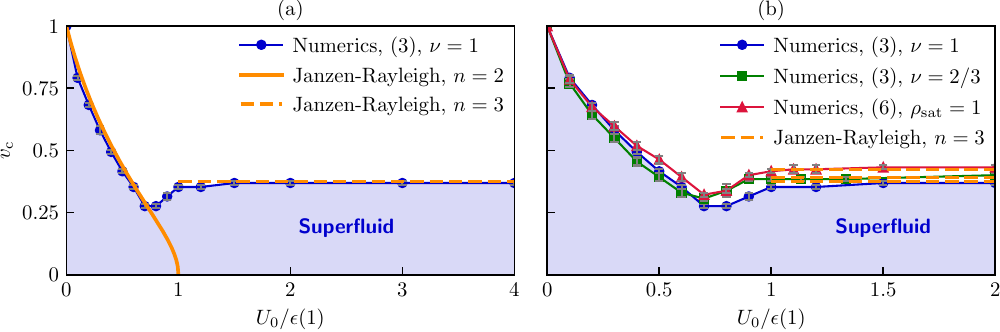}
\caption{(a) Critical velocity for superfluidity \maths{v_{\mathrm{c}}} as a function of \maths{U_{0}/\epsilon(1)} for an interaction potential \maths{\epsilon(\rho)} of the form~\eqref{Eq:MatterNL} with \maths{\nu=1}. We compare the numerical results (circles) with the analytical ones obtained using a Janzen-Rayleigh expansion of the velocity potential to the second order in the penetrable regime \maths{U_{0}/\epsilon(1)<1} (solid curve) and to the third order in the impenetrable regime \maths{U_{0}/\epsilon(1)>1} (dashed line). The numerical and analytical results accurately coincide at low and large \maths{U_{0}/\epsilon(1)} but differ around \maths{U_{0}/\epsilon(1)=1} (see main text). (b) Critical velocity for superfluidity \maths{v_{\mathrm{c}}} as a function of \maths{U_{0}/\epsilon(1)} for three different interaction potentials \maths{\epsilon(\rho)}:~\eqref{Eq:MatterNL} with \maths{\nu=1} [circles; same as in panel~(a)],~\eqref{Eq:MatterNL} with \maths{\nu=2/3} (squares), and~\eqref{Eq:LightNL} with \maths{\rho_{\mathrm{sat}}=1} (triangles). In all cases, we observe a  minimum of the critical velocity for \maths{U_{0}/\epsilon(1)<1}. The dashed lines represent the analytical critical velocities \maths{v_{\mathrm{c},3}} referenced in Tab.~\ref{Tab:VcImpenetrable}. For clarity, we do not show the analytical curves \maths{v_{\mathrm{c},2}=v_{\mathrm{c},2}(U_{0})} for \maths{U_{0}/\epsilon(1)<1}. In both panels, the vertical error bars correspond, at best, to the discretization of the velocity used in the numerical approach (see Appendix).}
\label{Fig:Vc}
\end{figure*}

We employ exactly the same method~\cite{Rica2001} as in Sec.~\ref{SubSec:PenetrableRegime} to solve Eqs.~\eqref{Eq:HydroSFHydrau},~\eqref{Eq:AC}, and~\eqref{Eq:BCImpenetrable}. Focusing on~\eqref{Eq:MatterNL} for instance, this yields the following results for the equivalents of Eqs.~\eqref{Eq:MatterVc0Penetrable},~\eqref{Eq:MatterJRPenetrable1}, and~(\ref{Eq:MatterJRPenetrable2},~\ref{Eq:MatterJRPenetrable3}), respectively:
\begin{equation}
\label{Eq:MatterVc0Impenetrable}
v_{\mathrm{c},0}=\sqrt{\frac{2}{8+3\nu}},
\end{equation}
\begin{align}
\notag
\nabla^{2}\phi_{k+1}&\left.=\frac{1}{2v_{\infty}^{2}}\sum_{j=0}^{k}\Big[\nabla\phi_{k-j}\cdot\nabla (v^{2})_{j}\right. \\
\label{Eq:MatterJRImpenetrable1}
&\left.\hphantom{=}+\nu\nabla^{2}\phi_{k-j}(v^{2})_{j}\Big]-\frac{\nu}{2}\nabla^{2}\phi_{k},\right.
\end{align}
\begin{equation}
\label{Eq:MatterJRImpenetrable2}
\frac{\partial\phi_{k}}{\partial r}\bigg|_{r=\sigma}=0.
\end{equation}
Note that when \maths{\nu=1}, one recovers the celebrated \maths{v_{\mathrm{c},0}=(2/11)^{1/2}=0.42(6)} first established in Ref.~\cite{Frisch1992} (first figure in Tab.~\ref{Tab:VcImpenetrable}). One should also notice the drastic difference in complexity between the Janzen-Rayleigh boundary conditions~\eqref{Eq:MatterJRImpenetrable2} recursively treated in Ref.~\cite{Rica2001} in the \maths{\nu=1} case, and the novel equations~(\ref{Eq:MatterJRPenetrable2}, \ref{Eq:MatterJRPenetrable3}) specific to the penetrable circular barrier and Eq.~\eqref{Eq:MatterNL} with an arbitrary \maths{\nu}. Table~\ref{Tab:VcImpenetrable} shows the accuracy of the Janzen-Rayleigh method in determining the critical speed for superfluid motion past the impenetrable barrier for the three interaction potentials of Fig.~\ref{Fig:Vc0}(b). In the \maths{\nu=1} case, one recovers the first values of the series established in Ref.~\cite{Rica2001}. In this impenetrable regime, the breakdown of superfluidity manifests by the nucleation of quantized vortices with opposite circulations at the north and south poles \maths{(r,\theta)=(\sigma,\pm\pi/2)} of the circular obstacle~\cite{Frisch1992, Pinsker2014, Kwak2023}.

\section{Comparison with numerical simulations and discussion}
\label{Sec:ComparisonWithNumericalSimulationsAndDiscussion}

In Ref.~\cite{Kwak2023}, the authors numerically investigate the critical velocity of a 2D nonlinear Schr\"odinger superflow past a Gaussian potential barrier, as described by Eq.~\eqref{Eq:GNLSE} with \maths{U(\boldsymbol{r})=U_{0}\exp(-r^{2}/\sigma^{2})} and \maths{\epsilon(\rho)=\rho}. They observe, for wide enough obstacle potentials (i.e., \maths{\sigma} large), that the critical velocity is a nonmonotonic function of \maths{U_{0}} which decreases for \maths{U_{0}/\epsilon(1)<1}, reaches a minimum around \maths{U_{0}/\epsilon(1)=1}, and slightly increases towards a \maths{U_{0}}-independent value for \maths{U_{0}/\epsilon(1)>1}. This is similar to what we have found analytically in Sec.~\ref{Sec:AnalyticalDerivationOfTheCriticalVelocity}, although the smallest value of the critical speed calculated in Ref.~\cite{Kwak2023} is not zero.

To bridge the gap between these numerical results and our analytical study done for the model potential barrier~\eqref{Eq:Obstacle}, we have performed similar numerical simulations in the case of a circular-shaped obstacle. We have used a finite-difference numerical scheme to determine the limit of the superfluid region where stationary solutions exist---all the technical details concerning the simulation method are given in the Appendix. To perform these simulations, we have used the smoothed-out circular potential barrier~\eqref{Eq:SmoothCylinder} with a radius \maths{\sigma=10} and a shoulder of width \maths{w=1}, which is different from the very wide circular potential~\eqref{Eq:Obstacle} of sharp boundary used in the analytical approach. Another difference is that the simulation takes into account the full Hamiltonian whereas the analytical theory neglects the dispersive term \maths{-\nabla^{2}\rho^{1/2}/(2\rho^{1/2})} in the hydrodynamic equations~\eqref{Eq:HydroSF}.

The comparison between our numerical and analytical results is shown in Fig.~\ref{Fig:Vc}(a) in the case of the interaction potential of Ref.~\cite{Kwak2023}, i.e., \maths{\epsilon(\rho)=\rho} [Eq.~\eqref{Eq:MatterNL} with \maths{\nu=1}]. The solid curve \maths{v_{\mathrm{c},2}=v_{\mathrm{c},2}(U_{0})} in the penetrable-barrier regime \maths{U_{0}/\epsilon(1)<1} is obtained using a Janzen-Rayleigh expansion of the velocity potential to the second order [i.e., Eq.~\eqref{Eq:JR} with \maths{n=2}] whereas the dashed line \maths{v_{\mathrm{c},3}=\mathrm{const}} in the impenetrable-barrier regime \maths{U_{0}/\epsilon(1)>1} is deduced from the same method but to the third order. The agreement between the numerical and analytical predictions is very good at both low and large \maths{U_{0}/\epsilon(1)}. This validates the existence of two distinct branches of solutions in the penetrable and impenetrable regimes, corresponding to a breakdown of superfluidity that happens inside or outside the obstacle through the nucleation of a rarefaction wave~\cite{Pinsker2014, Kwak2023} or of quantized vortices~\cite{Frisch1992, Pinsker2014, Kwak2023}, respectively. However, contrary to what the analytical approach predicts, these two branches turn out to be smoothly connected around the penetrable-to-impenetrable transition threshold \maths{U_{0}/\epsilon(1)=1} and the critical speed displays a nonzero minimum [\maths{\simeq0.28} in Fig.~\ref{Fig:Vc}(a)] in its vicinity, which is similar to what is observed in Ref.~\cite{Kwak2023}.

These smoothing and nonzero minimum are likely due to the dispersive term \maths{-\nabla^{2}\rho^{1/2}/(2\rho^{1/2})} omitted in the analytical approach but fully taken into account in the numerical simulations.\footnote{They probably also originate, but certainly in a more marginal way, from the continuous behavior of the potential barrier~\eqref{Eq:SmoothCylinder} around \maths{r=\sigma}.} Indeed, this term tends to reduce density gradients in the penetrable-barrier regime and then favors a larger density inside the obstacle. This yields a larger local speed of sound and then a larger and nonzero critical velocity close to the penetrable-to-impenetrable transition. We thus expect our two analytical curves, accurate away from \maths{U_{0}/\epsilon(1)=1}, to be smoothly connected around this threshold without vanishing. Next, the fact that the smoothing of the penetrable-to-impenetrable transition is accompanied with a minimum for the critical velocity can be understood as follows. During this transition, the points of emission of the excitations responsible for the breakdown of superfluidity are shifted from the interior of the obstacle towards the exterior (at the obstacle's boundary in our configuration) where the density is less depleted by the potential and then where the local speed of sound and so the critical velocity are larger. This explains the increase of the critical velocity at the transition and then the existence of a minimum for it right before, provided the mechanism responsible for the  breakdown of superfluidity shifts from the formation of a rarefaction wave inside the obstacle towards the nucleation of quantized vortices outside the obstacle.

In Fig.~\ref{Fig:Vc}(b), we compare the numerical results for the three interaction potentials \maths{\epsilon(\rho)} considered in Fig.~\ref{Fig:Vc0}(b), for the same circular potential barrier~\eqref{Eq:SmoothCylinder} with a radius \maths{\sigma=10} and a shoulder of width \maths{w=1}. The behavior is similar in all cases and a minimum for \maths{v_{\mathrm{c}}} is always observed for a ratio \maths{U_{0}/\epsilon(1)\sim0.7}. The minimal value of each critical velocity ranges from \maths{\simeq0.28} for~\eqref{Eq:MatterNL} with \maths{\nu=1}, to \maths{\simeq0.32} for~\eqref{Eq:LightNL} with \maths{\rho_{\mathrm{sat}}=1}. This is similar to what is observed for a Gaussian potential barrier \maths{U(\boldsymbol{r})=U_{0}\exp(-r^{2}/\sigma^{2})}, for which one obtains a critical velocity slightly lower going down to \maths{v_{\mathrm{c}}\sim0.2}~\cite{Kwak2023}. Finally, the numerical results compare favorably to the analytical ones at low and large \maths{U_{0}/\epsilon(1)}, as shown in Fig.~\ref{Fig:Vc}(b) but only in the impenetrable regime \maths{U_{0}/\epsilon(1)>1} for the sake of lisibility.

\section{Conclusion and outlook}
\label{Sec:ConclusionAndOutlook}

We have theoretically investigated the condition of existence of a 2D superflow past a static potential barrier of large width. Building atop Refs.~\cite{Frisch1992, Rica2001, Pinsker2014} (see also Refs.~\cite{Josserand1997, Josserand1999, Huepe2000, Stiessberger2000, Sasaki2010, Singh2017, Pigeon2021, Stockdale2021, Muller2022, Kwak2023, Ronning2023} for related theoretical work), we have analytically derived its critical velocity using the hydraulic approximation, the hodograph method, and Janzen-Rayleigh expansions of the velocity potential. This has been done in the case of an obstacle potential of certainly simplistic shape---we have focused on the circular potential barrier---but of \textit{arbitrary} penetrability, and in the case of a local interaction energy of \textit{arbitrary} dependence on the fluid density in the nonlinear Schr\"odinger equation. This pushes the state of the art towards more realistic modeling of recent experiments with atomic Bose-Einstein condensates~\cite{Kwon2015} and paraxial superfluids of light~\cite{Michel2018, Eloy2021} in 2D. Our analytical results have been shown to fairly agree with imaginary-time numerical calculations inspired from Ref.~\cite{Kwak2023}.

In strong contrast to the 1D geometry (see, e.g.,~\cite{Huynh2022} and references therein), the 2D critical velocity is a nonmonotonic function of the typical amplitude of the potential barrier~\cite{Kwon2015, Kwak2023}. When the obstacle potential is penetrable, the breakdown of superfluidity manifests by the emission of a rarefaction wave~\cite{Pinsker2014, Kwak2023} inside the potential and the corresponding critical speed decreases with the potential's amplitude. When the obstacle potential is on the contrary impenetrable, superfluidity breaks down at the potential's boundary where quantized vortices with opposite circulations~\cite{Frisch1992, Pinsker2014, Kwak2023} are nucleated at a critical speed independent of the potential's amplitude. These two different tendencies are smoothed out at the penetrable-to-impenetrable transition where the critical velocity displays a minimum.

An extension to the present work could be the study of the critical speed of a 2D superflow past a potential barrier narrower than the healing length. In this case, the density varies at the scale of the healing length (see, e.g., Ref.~\cite{Hakim1997} although in 1D) and the quantum pressure \maths{-\nabla^{2}\rho^{1/2}/(2\rho^{1/2})} can no longer be neglected, which has not yet been analytically tackled despite numerical investigations~\cite{Sasaki2010, Pinsker2014, Kwak2023}. Moreover, interaction between point-like impurities and a superfluid bath seems to be an important ingredient to understand nonequilibrium dynamics in recent experiments (see, e.g., Ref.~\cite{Cayla2023} although in 3D). As another prospect, it could be interesting to investigate the critical velocity resulting from the presence of several obstacles, or even in a disordered potential landscape where localization nontrivially competes with interactions (see, e.g.,~\cite{SanchezPalencia2010, Burmistrov2021} and references therein). In this case, the critical speed should become a random variable~\cite{Albert2010} whose statistical properties are unknown in 2D.

\begin{acknowledgments}
We acknowledge Amandine Aftalion, Tangui Aladjidi, Matthieu Bellec, Thomas Frisch, Quentin Glorieux, Maxime Ingremeau, Carelle Keyrouz, Claire Michel, Nicolas Pavloff, and Simon Pigeon for stimulating discussions and collaborations. P.-\'E. L. has appreciated the warm hospitality extended to him at Laboratoire Kastler Brossel where part of this study has been thought. This work has benefited from the financial support of Agence Nationale de la Recherche under Grants Nos.~ANR-21-CE30-0008 STLight and ANR-21-CE47-0009 Quantum-SOPHA.
\end{acknowledgments}

\appendix*

\section{Numerical calculation of the critical velocity}

Following the method proposed in Ref.~\cite{Kwak2023} for a similar problem with a Gaussian-shaped obstacle potential, we have used an imaginary-time numerical method to find superfluid stationary solutions when those exist.

In a reference frame where the obstacle is not moving, far away from the obstacle, a superfluid stationary solution of the dimensionless equation~\eqref{Eq:GNLSE} should behave as
\begin{equation}
\label{Eq:DefineVarphi}
\psi(\boldsymbol{r},t)=\exp\!\bigg\{i\bigg[v_{\infty}x-\bigg(\frac{v_{\infty}^{2}}{2}+\mu\bigg)t\bigg]\bigg\}=A(x,t)
\end{equation}
for a homogeneous fluid flowing in the \maths{x}-direction at constant velocity \maths{v_{\infty}>0}. We then look for solutions of the form \maths{\psi(\boldsymbol{r},t)=A(x,t)\varphi(\boldsymbol{r},t)} where the auxiliary function \maths{\varphi} should be independent of time \maths{t} and should tend towards \maths{1} far away from the obstacle if a stationary solution exists. Notice that \maths{\rho=|\psi|^{2}=|\varphi|^{2}}. We rewrite Eq.~\eqref{Eq:GNLSE} in terms of \maths{\varphi}, which gives the evolution equation \maths{i\partial\varphi/\partial t=\mathcal{H}\varphi}, where the Hamiltonian
\begin{equation}
\label{Eq:H}
\mathcal{H}=-\frac{1}{2}\nabla^{2}-iv_{\infty}\frac{\partial}{\partial x}+U(\boldsymbol{r})\mathbbm{1}_{U}+\epsilon(\rho)-\mu.
\end{equation}

Starting from an initial ansatz \maths{\varphi(\boldsymbol{r},0)}, we  propagate \maths{\varphi} in imaginary time \maths{\tau} using \maths{-\partial\varphi/\partial\tau=\mathcal{H}\varphi} and monitor if \maths{\varphi} converges towards a stationary solution at long \maths{\tau}. One should notice than no trivial imaginary-time dependence remains as the energy has been shifted to zero by the addition of the \maths{\mu}-term in Eq.~\eqref{Eq:H}. As shown in Ref.~\cite{Kwak2023}, when \maths{\varphi} does not tend towards a stationary solution, hydrodynamic perturbations that move perpendicularly to the flow of the superfluid are emitted around the obstacle. The observation of these emissions then shows the absence of a superfluid stationary solution, which can be monitored through the behavior of the rotational of the current, for example. On the contrary, in a regime where a stationary solution exists, we do not observe these perturbations and \maths{|\mathcal{H}\varphi|} decreases continuously towards zero.

To perform the imaginary-time evolution of \maths{\varphi}, we have used an explicit finite-difference scheme~\cite{Cerimele2000, Chiofalo2000, Minguzzi2004} which is conditionally stable for small enough time steps. We have not used a sharp circular potential as in Eq.~\eqref{Eq:Obstacle} but a smoothed version of it, of the form
\begin{equation}
\label{Eq:SmoothCylinder}
U(\boldsymbol{r})=\frac{U_{0}}{2}\bigg(1+\tanh\frac{\sigma-r}{w}\bigg),
\end{equation}
which is more amenable to our numerical scheme. Preliminary simulations have shown that a radius \maths{\sigma=10} is large enough to study the hydraulic approach introduced in Sec.~\ref{SubSec:HydraulicApproximation}. We have used a value of \maths{w=1} to avoid numerical difficulties, which means that we have not studied cases where the potential changes abruptly compared to the healing length \maths{\xi=1}. This is an important difference between the cases studied analytically and numerically.

We have typically used rectangular systems of sizes \maths{L_{x}=400} and \maths{L_{y}=100} with a space step \maths{\delta_{x}=0.25} and an imaginary-time step \maths{\delta_{\tau}=0.01}. As \maths{\varphi} goes towards a uniform solution \maths{\varphi=1} far away from the obstacle, we have used periodic boundary conditions. The wave function \maths{\psi} can be calculated from \maths{\varphi} using Eq.~\eqref{Eq:DefineVarphi} and used to evaluate quantities of interest such as currents. The linear size of the system along the \maths{x}-axis allows us to increment \maths{v_{\infty}} by steps \maths{\delta_{v}=2\pi/L_{x}\simeq0.016}. We have performed simulations to maximal imaginary times \maths{\tau_{\mathrm{max}}\sim10^{4}}. In this limit, the largest value of \maths{|\mathcal{H}\varphi|} in the rectangular grid is generally of order \maths{10^{-5}}. To validate our approach, we have reproduced the results obtained in Ref.~\cite{Kwak2023} for a Gaussian potential barrier \maths{U(\boldsymbol{r})=U_{0}\exp(-r^{2}/\sigma^{2})}.

For given values of \maths{U_{0}}, we have done simulations for different values of \maths{v_{\infty}} and observed if a stationary solution is reached or not. This has allowed us to determine a transition interval with a precision which is, at best, \maths{\delta_{v}}. One should notice that a sort of critical slowing-down happens close to the transition between the nonstationary and stationary regimes: As \maths{v_{\infty}} is decreased towards the transition value, the imaginary-time interval before the emission of some perturbation increases and this emission may not be observed in our finite-time window. We have defined the nonsuperfluid regime as the one where we observe the emission of several perturbations.

\bibliography{Vc2D}

\clearpage

\end{document}